\documentstyle[psfig,preprint,tighten,floats,aps]{revtex}

\def\B{S}
\def\M{{\cal M}}
\def\I{{\cal I}}
\def\P{{\cal P}}
\def\H{{\cal H}}

\def\i{\gamma}

\def\ut#1{\rlap{\lower1ex\hbox{$\sim$}}{#1}}

\def\l{\ell_{P}}
\def\C{{\rm C}}
\def\U{{\rm U}}
\def\SU{{\rm SU}}

\def\SL{{\rm SL}}
\def\E{{}^\i\!\Sigma}
\def\A{{}^\i\!\!A}
\def\ba{\begin{eqnarray}}
\def\ea{\end{eqnarray}}
\def\be{\begin{equation}}
\def\ee{\end{equation}}

\begin{document}
\draft
\title{Quantum Geometry and Black Hole Entropy}
\author{A.\ Ashtekar${}^1$, J.\ Baez${}^{2}$, A.\ Corichi${}^{1,3}$, 
K.\ Krasnov${}^1$}
\address{1. Center for Gravitational Physics and Geometry, \\
Pennsylvania State University, PA 16802, USA\\
2. Department of Mathematics, University of California, \\
Riverside, CA 92521, USA\\
3. Instituto de Ciencias Nucleares, UNAM \\
A. Postal 70-543, M\'exico D.F.\ 04510, M\'exico}

\maketitle

\begin{abstract}

A `black hole sector' of non-perturbative canonical quantum gravity is
introduced.  The quantum black hole degrees of freedom are shown to be
described by a Chern-Simons field theory on the horizon.  It is shown
that the entropy of a large non-rotating black hole is proportional to
its horizon area. The constant of proportionality depends upon the
Immirzi parameter, which fixes the spectrum of the area operator in loop
quantum gravity; an appropriate choice of this parameter gives the
Bekenstein-Hawking formula $S = A/4\l^2$.  With the {\em same choice} of
the Immirzi parameter, this result also holds for black holes carrying
electric or dilatonic charge, which are not necessarily near extremal.

\end{abstract}
\pacs{PACS: 04.60.-m, 04.70.Dy}

The statistical mechanical origin of black hole entropy has drawn a
great deal of attention recently (for reviews, see for example
\cite{Carlip,Horowitz,Rovelli}).  Most of the work based on string
theory has focused on the extremal and near extremal cases.  The
purpose of this letter is to introduce a new framework, based on
non-perturbative quantum gravity \cite{Ashtekar,RS}, that enables one
to treat general black holes in four dimensions.

The basic ideas can be summarized as follows.  We first introduce a
sector of the classical theory that corresponds to isolated,
non-rotating black holes and find the associated phase space
description.  Then we quantize the resulting phase space.  Finally, we
isolate the quantum states that describe the geometry of the
horizon. It is these degrees of freedom that account for the black
hole entropy in our approach. We find that the statistical mechanical
entropy of the black hole is proportional to its horizon area.

Recently, non-perturbative techniques have led to a quantum theory of
geometry in which operators corresponding to lengths, areas and
volumes have discrete spectra.  Of particular interest are the spin
network states associated with graphs in 3-space with edges labelled
by spins $j = {1\over 2},1, \dots$ and vertices labelled by
intertwining operators \cite{alb,network}. If a single edge punctures
a 2-surface transversely, it contributes an area proportional to
$\sqrt{j(j+1)}$ \cite{area,al}.  Over the last two years, this picture
led to certain constructions which in turn inspired the present
work. First, while working with a space-time region with boundary in
Euclidean general relativity with non-zero cosmological constant,
Smolin \cite{Smolin} was led to introduce gravitational surface states
which could be identified with the states of the $\SU(2)$ Chern-Simons
theory on a surface with punctures.  Second, Rovelli \cite{cr},
motivated by the work of Krasnov \cite{GeomEntr}, estimated the number
of spin-network states which endow a 2-sphere with a given, large area
and applied this estimate to black hole horizons.  Third, Krasnov
\cite{K} proposed to combine the two sets of ideas by introducing
certain boundary conditions on regions bounded by 2-spheres. Finally,
Carlip's \cite{Carlip} considerations of surface states in the context
of three dimensional black holes also played a suggestive role in the
classical part of our treatment.

Let us begin with uncharged, non-rotating black holes. A sector of the
classical phase space corresponding to an isolated, non-rotating black
hole can be constructed as follows. Consider the manifold (with
boundary) representing the asymptotic region of FIG. 1. We refer to
the outer boundary as $\I$ and the inner boundary as $\H$.  Our
dynamical fields are a soldering form $\sigma_a^{AA'}$ for $\SL(2,\C)$
spinors and an $\SL(2,\C)$ connection $A_{aA}{}^{B}$
\cite{Ashtekar,4}. (In a classical solution, $g_{ab} = \sigma_a^{AA'}
\sigma_{bAA'}$ is the Lorentzian space-time metric and $A_{aA}{}^B$ is
the self-dual connection that operates only on unprimed spinors.) On
$\I$, fields are required to satisfy the standard asymptotically flat
boundary conditions. The conditions on $\H$, on the other hand, are
more subtle and will be discussed in detail elsewhere. The key
requirements are: i) $\H$ be a null surface with respect to the metric
$g_{ab}$; ii) On a `finite patch' $\Delta$ of $\H$, the area of any
cross-section be a constant, $A_{\B}$, the Weyl spinor be of Petrov
type 2-2 and its only non-zero component, $\Psi_2$, be given by
$\Psi_2 = {2\pi}/A_{\B}$; and, iii) the 2-flats on $\Delta$,
orthogonal to the two principal null directions of the Weyl tensor
span 2-spheres and the pull-back of the connection $A_a$ to these
2-spheres be real.

In what follows, we focus only on the part $\Delta$ of $\H$ and the
corresponding region $\cal M$ of the spacetime (see FIG. 1). Roughly,
condition ii) implies that there is no gravitational radiation falling
into $\Delta$ (i.e., the black hole is `isolated' there) while the
first part of iii) implies that it is `non-rotating'. The three
conditions together imply that, on partial Cauchy surfaces (such as
$M$) that intersect $\Delta$ in the preferred 2-spheres, the
2-spheres are marginally outer trapped surfaces. These boundary
conditions have been extracted from the geometrical structure
available at the Schwarzschild horizon. However, we do {\it not}
require staticity and allow gravitational waves in the exterior
region. Therefore, our phase space will be infinite dimensional; the
boundary conditions are quite weak.

\begin{figure}[h]
\centerline{\hbox{
\psfig{figure=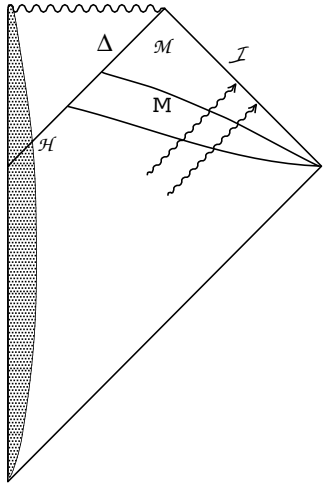,height=2.5in}}}
\caption{Example of a spacetime of interest.}
\end{figure}

However, these boundary conditions  are strong enough to imply
that the variational principle is well defined.  More precisely, we
can add to the standard self-dual action \cite{4} a surface term so
that the total action is functionally differentiable and yields precisely
Einstein equations:
\begin{eqnarray} \label{action}
S(\sigma,A) = -\,{i\over 8\pi G}\int_\M\,{\rm Tr}\,
\left(\Sigma\wedge F\right)\nonumber \\
- {i\over 8\pi G}{A_{\B}\over 4\pi}\int_{\Delta}{\rm Tr}
\left(A\wedge dA + {2\over 3}A\wedge A\wedge A\right).
\end{eqnarray}
Here $\Sigma_{ab}^{AB} = 2 \sigma_{[a}{}^{AA'}\sigma_{b]A'}{}^B$ while
$F_{abA}{}^B$ is the curvature of the connection $A$, and $G$ is
Newton's constant. (Throughout we have set $c = \hbar =1$.)  Note that
the required surface term is precisely the action of Chern-Simons
theory. It is straightforward to cast the theory into Hamiltonian
form.  The basic phase space variables are the restrictions of
$\Sigma$ and $A$ to the spatial hypersurface $M$ with a boundary
$\B$. (Vector densities, dual to the pull-back of $\Sigma$, are the
familiar density weighted triads.) Unfortunately, the restriction to
$M$ of the self-dual connection $A$ is a complex valued $\SU(2)$
connection and the functional analysis required to handle complex
connections in quantum theory is not yet fully developed.

Therefore, at this stage, it is easier to make a transformation to
real variables \cite{5}. On $M$, $A$ can be expressed in terms of real
fields as $A_a = \Gamma_a - i K_a$, where $\Gamma$ is the
3-dimensional spin connection compatible with the triad field and $K$
is the extrinsic curvature of $M$. This suggests \cite{ImmPar} that we
introduce real phase space variables $\A_a := \Gamma_a - \i K_a$ and
$\E_{ab} := (1/\i) \Sigma_{ab}$, where $\i$ is a positive real number
known as Immirzi parameter. Then, our boundary conditions imply
that the phase space consists of real fields $(\A_a, \E_{ab})$ that
are asymptotically flat at infinity and satisfy the following
condition at $\B$:
\be \label{bc}
{}^\gamma\!\underline{F}_{ab}^{AB} = - \frac{2\pi\gamma}{A_{\B}}\,\, 
{}^\gamma\!\underline{\Sigma}_{ab}^{AB} \, ,
\ee
where underbars denote pull-backs to $\B$. This in turn implies that
the restriction of $\A_a$ to $\B$ yields a reducible connection, i.e.,
one satisfying $D_a r = 0$ for some `radial' internal vector
$r$. Although it is not necessary to do so, for technical simplicity,
we fix $r$ on $\B$ using the $\SU(2)$ gauge freedom. Then the
gauge group on the boundary is reduced to $\U(1)$ and only the $r$
component of (\ref{bc}) is non-trivial.
  
On this manifestly real phase space, the symplectic structure 
derived from (\ref{action}) is given by
\ba 
\Omega|_{(\A,\E)} 
\left ( (\delta\A,\delta\E), (\delta\A',\delta\E') \right ) =
\nonumber \\
{1\over 8\pi G} \int_M {\rm Tr}
\left [ \delta\E\wedge\delta\A' - \delta\E'\wedge\delta\A \right ]
\label{ss} \\
- \frac{k}{2\pi} \oint_\B  \,{\rm Tr} 
\left [ \delta\A\wedge\delta\A' \right ],
\nonumber
\ea
\begin{equation}
{\rm where}\quad k = {A_{\B}\over 8\pi\gamma G}
\label{level}
\end{equation}
is later identified with the level of the Chern-Simons theory.
Up to a numerical coefficient, $k$ is simply the area of the horizon
of black hole measured in the units of Planck area $\l^2 = G$
\cite{K}.  Note that, in addition to the familiar volume term, the
symplectic structure has a surface term which coincides with the
symplectic structure of the Chern-Simons theory.

The theory has three sets of first class constraints.  A careful
analysis shows that they generate the following gauge transformations:
i) $\SU(2)$ internal rotations that reduce to $\U(1)$ rotations
preserving a fixed vector $r$ on the boundary $\B$; ii) spatial
diffeomorphisms that leave $\B$ invariant; and, iii) canonical
transformations generated by the scalar constraint with lapse fields
approaching zero at spatial infinity and on $\B$.  Somewhat
surprisingly, it turns out that condition (\ref{bc}), the pull-back to
$\B$ of the type 2-2 requirement, ensures full gauge invariance on the
boundary.  Without it, as in the case of the scalar constraint, only
the internal rotations whose generators vanish on $\B$ could be
regarded as gauge.

It is intuitively clear that not all the degrees of freedom described
by fields $\A, \E$ are relevant to the problem of black hole entropy.
In particular, there are `volume' degrees of freedom in the theory
corresponding to gravitational waves far away from $\Delta$ which
should not be taken into account as genuine black hole degrees of
freedom.  The `surface' degrees of freedom describing the geometry of
the horizon $\B$ have a different status.  It has often been argued
(see, e.g., \cite{Rovelli} and references therein) that it is the
degrees of freedom `living on the horizon' that should account for the
entropy.  We adopt this viewpoint in our approach.

In the classical theory that we have described, the volume
and surface degrees of freedom cannot be separated: all fields on $\B$
are determined by fields in the interior of $M$ by continuity.
However, in the quantum theory, the fields describing geometry become
discontinuous in certain precise sense \cite{alb}, and the fields on
$\B$ are no longer determined by fields in $M$; in this case there are
independent degrees of freedom `living' on the boundary.  These
surface degrees of freedom are the ones that account for black hole
entropy in our approach.

To quantize the theory, we first construct a Hilbert space $\H^V$ of
`volume' states and a Hilbert space $\H^S$ of `surface' states, and
then impose constraints on $\H^V \otimes \H^S$ to obtain the space of
physical states.  We take $\H^V$ to consist of certain
square-integrable functions on the space of generalized $\SU(2)$
connections \cite{alb} on $M$ modulo gauge transformations that are
the identity on $\B$.  The form of the Hilbert space $\H^S$ of surface
states is motivated by the fact that in the quantum theory we wish to
impose the boundary condition (\ref{bc}) as an operator equation.
That is, given a spin network state $\Psi_V$ in $\H^V$ and a state
$\Psi_S$ in $\H^S$, the quantum version of the $r$ component of
equation (\ref{bc}) should read

\be\label{qbc} 
(1{\textstyle\bigotimes} {2\pi\gamma\over A_{\B}}\hat{
\underline{F}}_{ab}\cdot r
+ \hat{\underline{\Sigma}}_{ab} \cdot r
{\textstyle\bigotimes} 1)\, \Psi_V {\textstyle\bigotimes} \Psi_\B\, 
= 0 \ee
The structure of this equation implies that $\Psi_V$ and $\Psi_B$
should be eigenstates of $\hat{\underline{\Sigma}}_{ab}\cdot r$ and
$\hat{\underline{F}}_{ab}\cdot r$ respectively. Now, the `polymer
nature' of quantum geometry in $M$ implies that eigenvalues of
${\hat{\underline\Sigma}_{ab}\cdot r}$ are distributional, given by
\cite{al}
\be
8\pi \l^2\sum_{i}\, j_i \delta^2(x,p_i) \eta_{ab}\, \l^2
\ee
for some points $p_i$ on $\B$, where $j_i$ are half-integers,
$\delta^2$ is the delta distribution on $\B$, $\eta_{ab}$ the
Levi-Civita density on $\B$ and $\l$ the Planck length. Therefore,
(\ref{qbc}) implies that the surface states $\Psi_S$ have support only
on generalized connections that are everywhere flat except at a
finite number of points $p_i$. It turns out that such generalized
connections can be identified with ordinary connections with
distributional curvature.  Since the surface symplectic structure is
that of Chern-Simons theory, for any fixed choice
$$\P=\{(p_1,j_{p_1}),\ldots,(p_n,j_{p_n})\}$$
of points in $\B$ labelled by spins, we wish $\H^S$ to have a subspace
given by the space of states of $\U(1)$ Chern-Simons theory on a
sphere with punctures $p$ labelled by spins $j_p$. The total
space $\H^S$ is the direct sum of these subspaces.

Note now that $(k/2\pi) \hat{\underline F}$ is the generator of
internal rotations in Chern-Simons theory. Thus, the meaning of
(\ref{qbc}) turns out to be rather simple: it ensures that the volume
and surface states are `coupled' in precisely the correct way so that
the total state is invariant under $\U(1)$ internal rotations at $\B$.
The remaining constraints require that the states be invariant under
diffeomorphisms of $M$ that leave $\B$ invariant and under motions
generated by the Hamiltonian constraint smeared with any lapse field
that vanishes at $\B$. Thus, the following physical picture
emerges. For each set $\P$ of finitely many punctures $p$ labelled by
spins $j_{p}$, there is a subspace $H^V_\P$ of volume states having a
basis given by open spin networks whose edges intersect $\B$ only at
these punctures, where they are labeled by the spins $j_{p}$.
Similarly there is a subspace $H^\B_\P$ consisting of quantum states
of $\U(1)$ Chern-Simons theory on the punctured surface $\B$.  The
total physical Hilbert space is given by:
$$\H_{\rm phy} = 
{\bigoplus_\P\, \left [\H^V_\P \otimes \H^S_\P \right ]
\over{\rm Gauge}},$$ 
where `Gauge' means internal $\SU(2)$ rotations that reduce to $\U(1)$
on $\B$, diffeomorphisms preserving $\B$, and the motions generated by
the Hamiltonian constraint. The quotient by diffeomorphisms identifies
any two Hilbert spaces associated with sets $\P$ that can be mapped
into another by a diffeomorphism on $\B$.  Thus, what matters is only
the spins labelling punctures, not the locations of individual
punctures.  Unfortunately, we do not have yet a complete control over
the quantum Hamiltonian constraint, despite the recent progress on
this front \cite{Dynamics}.  To proceed, we make a rather weak
assumption about the quantum dynamics: namely, that generically there
is at least one solution of this constraint in $\H^V_P \otimes
\H^S_\P$ for any set $\P$ of punctures labelled by spins.

We are not interested in this full Hilbert space since it includes,
e.g., states of gravitational waves far away from $\Delta$.  Rather,
we wish to consider only states of the horizon of a black hole with
area $A$.  Thus we trace over the `volume' states to construct an
density matrix $\rho_{\rm bh}$ describing a maximal-entropy mixture of
surface states for which the area of the horizon lies in the range $A_{\B}
\pm \l^2$.  The statistical mechanical black hole entropy is then
given by $S_{\rm bh} = - {\rm Tr} \rho_{\rm bh} \ln \rho_{\rm bh}$.
As usual, this can be computed simply by counting states: $S_{\rm bh}
= \ln N_{\rm bh}$ where $N_{\rm bh}$ is the number of Chern-Simons
surface states satisfying the area constraint.

Fortunately, the eigenvalues of the area operator are explicitly known. 
For the case now under consideration, they are given by \cite{area}: 
\be \label{ev} 
8\pi\i \l^2 \sum_p \sqrt{j_p(j_p+1)} 
\ee
where $j_p$ are the spins labelling the punctures. Using this and the
fact (from Chern Simons theory) that for a large number of punctures
the dimension of $\H^S_\P$ grows as
\begin{equation}
{\rm dim} \H^S_\P \,\, \sim \,\, \prod_{j_p\in\P} (2j_p+1) ,
\end{equation}
it is straightforward to calculate the entropy.  For large $A$ it is
given by 
\begin{eqnarray}\label{ent}
S_{\rm bh} = \frac{\i_0}{4\l^2\i}\,  A_{\B} ,\quad
\i_0 = \frac{\ln{2}}{\pi\sqrt{3}},
\nonumber
\end{eqnarray}
where the appearance of $\i$ can be traced back directly to the formula
for the eigenvalues of the area operator, (\ref{ev}).  Thus, in the
limit of large area, the entropy is proportional to the area of the
horizon.  If we set $\i= \i_0$, the statistical mechanical entropy is given
precisely by the Bekenstein-Hawking formula.  

Are there independent checks on this preferred value?  The answer is
in the affirmative.  One can carry out this calculation for
Reissner-Nordstrom as well as dilatonic black holes.  A priori it
could have happened that, to obtain the Bekenstein-Hawking value, one
would have to re-adjust the Immirzi parameter for each value of the
electric or dilatonic charge. This does {\em not} happen.  The entropy
is still given by (\ref{ent}) and hence by the Bekenstein-Hawking
value when $\i =\i_0$.

To summarize, we first introduced a black hole sector of the
gravitational phase space and then quantized it using the by now
well-developed framework of non-perturbative quantum gravity together
with results from Chern-Simons theory.  We found that the entropy is
proportional to the area irrespective of the value of the Immirzi
parameter $\i$ and that a single choice of $\i$ yields the
Bekenstein-Hawking coefficient irrespective of the parameters labeling
the non-rotating black hole.

We conclude with a few remarks:\\
i) One can show \cite{ImmPar} that different values of $\i$ correspond
to different `sectors', that is, unitarily inequivalent
representations of the canonical commutation relations.  The spectrum
of the area operator is different in each representation.  As usual in
such situations, the `correct' sector can only be singled out by
additional input (see, for example, the analogous ambiguity in the
loop quantization of Maxwell theory \cite{9}).  The Bekenstein-Hawking
calculation can be regarded as serving this purpose. However, the full
significance of $\gamma$ is yet to be understood.\\
ii) A detailed calculation shows that the states which dominate the
counting correspond to punctures all of which have labels $j=
1/2$. Thus, there is a curious similarity between our detailed results
and John Wheeler's ``It from Bit'' picture \cite{Wheeler} of the 
origin of black hole entropy.\\ 
iii) So far, we have only considered non-rotating black
holes. However, the basic ideas underlying this framework apply also
to the rotating case. \\
iv) Our approach provides only an `effective' description of a quantum
black hole, for we first isolated a black hole sector classically and
then quantized that sector.  The issue of extracting this sector from a
complete theory of quantum gravity is yet to be explored.  Nonetheless,
it is rather striking that subtle results from quite different areas ---
classical general relativity, quantum geometry and Chern-Simons theory
--- fit tightly without a mismatch to provide a coherent picture of the
microstates of a black hole.  The detailed implications of this picture
for the black hole evaporation process are now being explored \cite{10}.

{\bf Acknowledgements}: We are grateful to Don Marolf and Carlo Rovelli
for their comments. This work was supported in part by the NSF grant
PHY95-14240 and by the Eberly research funds of Penn State.  AA, JB
and KK acknowledge support from the Erwin Schr\"odinger Institute for
Mathematical Sciences.  AC was supported by DGAPA of UNAM.

\end{document}